\documentclass[prb,twocolumn,groupedaddress,showpacs]{revtex4}
\usepackage{graphicx}
\usepackage{latexsym}
\usepackage{amssymb}
\usepackage{amsfonts}
\begin{document}
\bibliographystyle{apsrev}
\title{Phase diagram of a superconductor / ferromagnet bilayer}
\author{M. Lange}
\email{martin.lange@fys.kuleuven.ac.be}
\author{M. J. Van Bael}
\author{V. V. Moshchalkov}
\affiliation{Laboratory for Solid State Physics and Magnetism, Nanoscale
Superconductivity and Magnetism Group, K.U. Leuven, Celestijnenlaan 200 D, 3001
Leuven, Belgium}
\date{\today}
\begin{abstract}
The magnetic field ($H$) - temperature ($T$) phase diagram of a superconductor is
significantly altered when domains are present in an underlying ferromagnet with
perpendicular magnetic anisotropy. When the domains have a band-like shape, the
critical temperature $T_c$ of the superconductor in zero field is strongly reduced,
and the slope of the upper critical field as a function of $T$ is increased by a
factor of $2.4$ due to the inhomogeneous stray fields of the domains. Field
compensation effects can cause an asymmetric phase boundary with respect to $H$ when
the ferromagnet contains bubble domains. For a very inhomogeneous domain structure,
$T_c \propto H^2$ for low $H$ and $T_c \propto H$ for higher fields, indicating a
dimensional crossover from a one-dimensional network-like to a two-dimensional
behavior in the nucleation of superconductivity.
\end{abstract}

\maketitle
\section{Introduction}
In hybrid superconductor / ferromagnet (SC/FM) bilayers the FM modifies quite
substantially the superconducting properties of the SC layer. In particular, strong
vortex pinning was reported recently for superconducting films covering arrays of
ferromagnetic dots with in-plane \cite{otani93, martin97, vanbael99} and out-of-plane
magnetization, \cite{morgan98, vanbael00} and for continuous SC/FM bilayers.
\cite{bula00a, garcia00, zhang01, besp01b, lange02} Theoretical investigations showed
that supercurrents and vortices can be induced in the SC by the stray field of the FM,
\cite{marmorkos96, milosevic02, lyu98, erdin02a, besp01a, sonin} and that the domain
structure of soft FM's can be influenced by the presence of the SC. \cite{buzdin88}\\
Furthermore, \citeauthor{radovic91} predicted the appearance of the so-called
$\pi$-phase state in SC/FM multilayers, where the phase of the superconducting order
parameter $\psi$ shifts by $\pi$ when crossing a ferromagnetic layer. \cite{radovic91}
Recently the existence of the $\pi$-phase state was confirmed by observing sharp cusps
in the temperature dependence of the critical current in SC/FM/SC junctions.
\cite{ryazanov01} Earlier experiments were performed in order to find the $\pi$-phase
state by measuring the predicted oscillatory dependence of the critical temperature
$T_{c}$ of SC/FM multilayers on the FM layer thickness $d_{fm}$. \cite{jiang95,
muhge96, aarts97, lazar00} However, the results of these experiments were not
conclusive, because the nonmonotonic $T_{c}(d_{fm})$ behavior could also appear due to
the presence of magnetically "dead" layers at the SC/FM interfaces. \cite{muhge96}\\
The theory of the anomalous $T_{c}(d_{fm})$ dependence is based on the Usadel
equations \cite{usadel70} describing the proximity effect of FM and SC layers, but
neglecting a possible influence of the domains in the FM. In this manuscript we will
show that an inhomogeneous stray field $B_{stray}$ produced by the domain structure of
a FM can actually also lead to a significant change in $T_{c}$. To demonstrate this
effect, we measure $T_{c}$ as a function of the perpendicularly applied magnetic field
$H$ of a Pb film on top of a Co/Pt multilayer with perpendicular magnetic anisotropy.
In this sample the proximity effect is suppressed by an amorphous Ge layer between Pb
and Co/Pt. The domain structure in the Co/Pt multilayer can consist of stable band or
bubble domains. The FM layer can also be in a single domain state, depending on the
preceding magnetization procedure, as was shown in a recent study of the vortex
pinning in this system. \cite{lange02} Due to field cancellation effects between $H$
and $B_{stray}$, and due to the suppression of $\psi$ by $B_{stray}$, $T_{c}(H)$ can
be controlled by changing the microscopic domain
structure.\\
\section{Magnetic properties of the C\lowercase{o}/P\lowercase{t} multilayer}
\label{secmag} 
\begin{figure}
\includegraphics{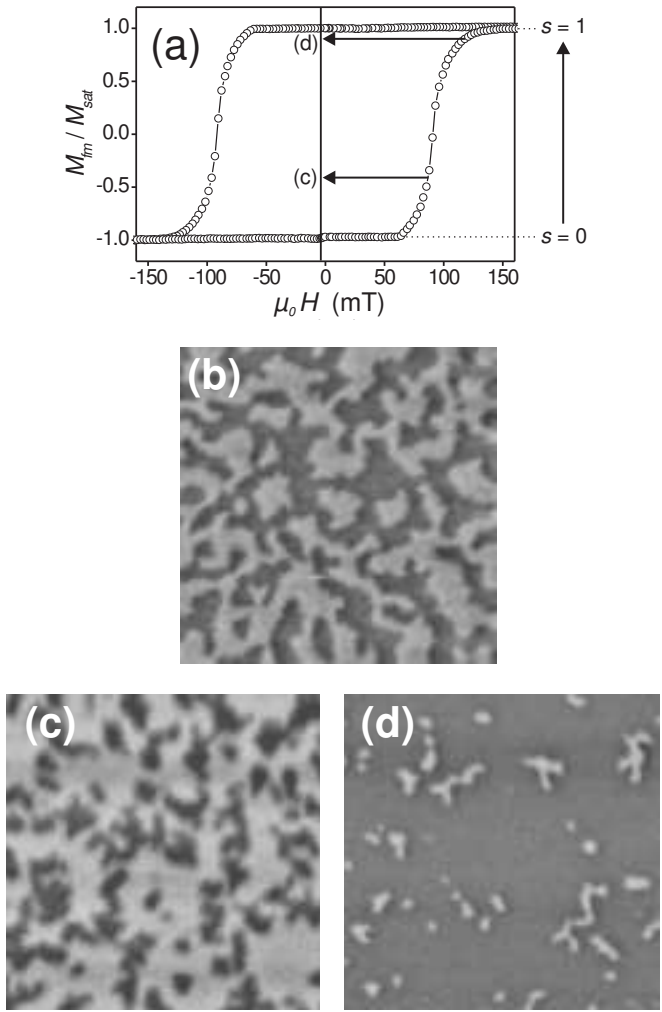}
\caption{Magnetic properties of the Co/Pt multilayer: (a)~Hysteresis loop measured by
magneto-optical Kerr effect with $H$ perpendicular to the sample surface. MFM images
(5~$\times$~5~$\mu\mathrm{m}^{2}$) show that the domain structure of the sample
consists of band domains after out-of-plane demagnetization (b), bubble domains in the
$s=0.3$ (c) and $s=0.93$ (d) states.}
 \label{moke}
\end{figure}

The properties of the Co/Pt multilayer have been described before.
\cite{lange02} Briefly, the multilayer has a [Co(0.4~nm)/Pt(1.0~nm)]$_{10}$ structure
on a 2.8~nm Pt base layer on a Si/SiO$_{2}$ substrate. The magnetic properties were
characterized by the magneto-optical Kerr effect (MOKE) and magnetic force microscopy
(MFM), revealing that the sample has perpendicular magnetic anisotropy.
Fig.~\ref{moke} shows the magnetization $M_{fm}$ of the Co/Pt multilayer measured by
MOKE, normalized to the saturation magnetization $M_{sat}$, as a function of the
magnetic field $H$ applied perpendicular to the surface.
The loop has an almost rectangular shape with $\mu_{0} H_{n} = 60$~mT, $\mu_{0} H_{c}
= 93$~mT, and $\mu_{0} H_{s} = 145$~mT, where $H_{n}$, $H_{c}$ and $H_{s}$ are the
nucleation, coercive and saturation field, respectively, and $\mu_{0}$ is the
permeability of the vacuum.\\ Using different magnetization procedures, one can
produce different stable domain patterns in the sample. For instance, after
out-of-plane demagnetization, band domains are observed by MFM, see
Fig.~\ref{moke}(b). Stable bubble domains with local magnetic moments $\textbf{m}$
either pointing up ($m_z>0$) or down ($m_z<0$) perpendicular to the sample surface can
be created by applying a negative field of $-1$~T, sweeping $H$ to a positive value
between $H_{n}$ and $H_{s}$, and then removing $H$. The parameter $s$, which gives the
fraction of magnetic moments that are pointing up ($m_z>0$) to the total amount of
magnetic moments, is used to describe the different remanent magnetic states obtained
after this magnetization procedure. The value of $s$ can be found from the MFM images
by dividing the dark area ($m_z>0$) by the total area, or from measurements of
$M_{fm}$ as will be described later.\\ The lateral size of the domain structures can
be estimated from the MFM images. The typical diameter of the bubble domains is about
$\sim 300$~nm. The same value is obtained for the average width of the band domains.
Although the magnetic moments of the Co/Pt multilayer are equally distributed between
up- and down-directions in both the demagnetized and the $s=0.5$ state, there are
distinct differences between these two domain states: In the demagnetized state the
lateral size of the domain is larger (because band domains are extended in one
direction). Note also that in the demagnetized state, the boundary between domains
with magnetization pointing up and down is well defined, but not straight: several
sharp corners of different angles can be seen. No MFM images could be obtained for the
$s=0.5$ state, caused by the difficult magnetization procedure due to the steep slope
of the $M_{fm}(H)$ curve, see Fig.~\ref{moke}(a). However, from the image of the
$s=0.3$ state, see Fig.~\ref{moke}(c), one can observe that the domain walls are less
sharp defined than in the demagnetized state.
\section{Phase boundary of the superconducting film}
\begin{figure}
\includegraphics{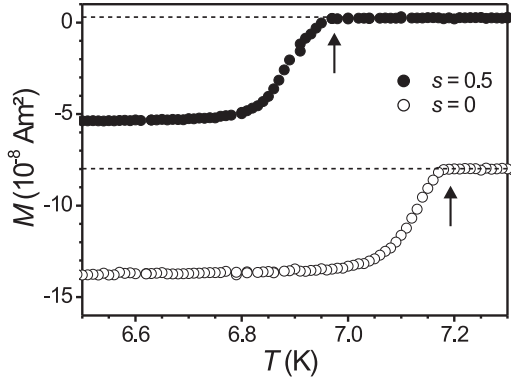}
\caption{Field cooled $M(T)$ measurements of the hybrid SC/FM bilayer in $\mu_{0} H =
0.5$~mT with the Co/Pt multilayer in the $s=0.5$ ($\bullet$) and $s=0$ ($\circ$)
states. The arrows indicate $T_{c}$.}
 \label{mt}
\end{figure}

After characterizing the properties of the FM, a 10~nm Ge film, a 50~nm Pb film and a
30~nm Ge capping layer are subsequently evaporated on the Co/Pt multilayer at a
substrate temperature of 77~K. The amorphous Ge film between Pb and Co/Pt is
insulating at low temperatures, so that the proximity effects between Pb and Co/Pt are
suppressed.\\ The upper critical field $H_{c2}$ of bulk type-II SCs is given by
\cite{tinkham}
\begin{eqnarray}
\mu_{0} H_{c2}(T) = \frac{\Phi_{0}}{2 \pi \xi^{2}(T)} \label{one}.
\end{eqnarray}
with $\Phi_{0}$ = 2.068~mT~$\mu$m$^{2}$ the superconducting flux quantum, $\xi(T) =
\xi(0)/\sqrt{1-T/T_{c0}}$ the temperature dependent coherence length in the dirty
limit, and $T_{c0}$ the critical temperature at zero field. Hence, the linear slope of
$H_{c2}$ as a function of temperature is only determined by the coherence length
$\xi$.\\ $H_{c2}(T)$ can behave differently when the geometry of the SC is changed,
e.g. for thin films with thickness $w<\xi(T)$. While eq.~\ref{one} is still valid for
thin type-II superconducting films with $H$ applied perpendicular to the sample
surface, $T_{c}$ for parallel $H$ is given by \cite{tinkham63}
\begin{eqnarray}
T_{c}(H) = T_{c0}[1 - \frac{\pi^{2} \xi^{2}(0) w^{2}}{3 \Phi_{0}^{2}} \mu_{0}^{2}
H^{2}] \label{two},
\end{eqnarray}
with $T_{c0}$ the zero-field critical temperature. In fact, this formula also gives
the phase boundary of a mesoscopic line in perpendicular field, because the cross
section, exposed to the applied field, is the same for a film of thickness $w$ in
parallel $H$ and for a mesoscopic line in perpendicular $H$. \cite{vvm95} For multiply
connected mesoscopic lines, $T_{c}(H)$ can show an even more complex behavior due to
fluxoid quantization effects. \cite{bruyndoncx96}\\ The phase boundary of the SC/FM
bilayer was measured in a Quantum Design superconducting quantum interference device
(SQUID) magnetometer with $H$ applied perpendicular to the surface. Fig.~\ref{mt}
shows the data obtained in two field cooled measurements of the total magnetization $M
= M_{fm} + M_{sc}$ ($M_{sc}$ is the magnetization of the SC) at the applied field of
$\mu_{0} H = 0.5$~mT, after the samples were brought in the $s=0.5$ and $s=0$ states.
Above $T_{c}$, $M$ has a constant value for both states, given by the contribution of
the FM $M_{fm}$, from which $s$ can be derived. When the sample is cooled through
$T_{c}$, a diamagnetic response of the SC appears. These kinds of measurements were
used to determine $T_{c}(H)$ as the temperature where $M$ starts to deviate from
$M_{fm}$. Repeating these measurements at several applied fields $|H|<25$~mT did not
change the offset $M_{fm}$ above $T_{c}$, implying an unchanged domain state.

\subsection{$T_c(H)$ with magnetized FM}
\begin{figure}
\includegraphics{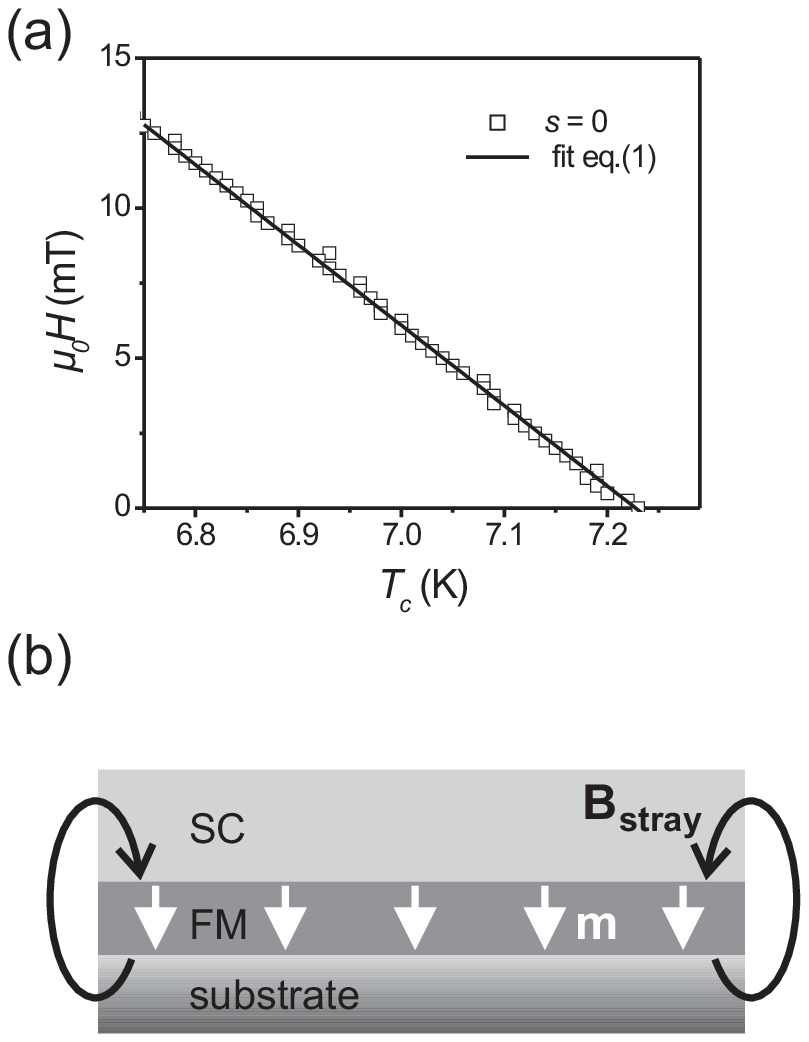}
\caption{(a)~Magnetic field - temperature phase diagram of the superconducting Pb film
covering the Co/Pt multilayer. Before measuring $T_{c}(H)$, the Co/Pt multilayer was
brought into the $s=0$ state. (b)~Schematic drawing of the stray field $B_{stray}$
when the FM is in the $s=0$ state.}
 \label{pbmag}
\end{figure}

The phase boundary for the $s=0$ state (all $m_z < 0$) obtained by several $M(T)$
measurements in varying fields is shown in Fig.~\ref{pbmag}(a).
A linear behavior of the phase boundary is observed, which can be fitted by
eq.~\ref{one} with $\xi(0)=(41.2 \pm 0.2)$~nm and $T_{c0}=(7.227 \pm 0.002)$~K. This
implies that in this state, the FM has no influence on the superconducting film,
because both the linear behavior and the values of $T_{c0}$ and $\xi(0)$ are in good
agreement with those of pure Pb films. \cite{lievethesis} It is important to note that
the temperature dependence of $\xi (T) = \xi (0) / \sqrt{1-T/T_{c0}}$ derived for this
domain state is the same for all domain states, since we are always dealing with the
same Pb film. \\ Let us consider the magnetic stray field $B_{stray}$ of a
homogeneously magnetized film in the $s=0$ state, schematically drawn in
Fig.~\ref{pbmag}(b). $B_{stray}$ has its largest amplitude at the sample boundary and
is negligible above the center of the FM. Intuitively this can be understood by
considering the stray field of a single magnetic dipole in the center of the sample:
The negative field above the dipole is compensated by the returning positive stray
field of the surrounding magnetic dipoles. Therefore, the main central part of the
superconductor is only weakly influenced by $B_{stray}$, and the measured $T_c(H)$
curve resembles the one of a single Pb film.

\subsection{$T_c(H)$ with demagnetized FM}
\begin{figure}
\includegraphics{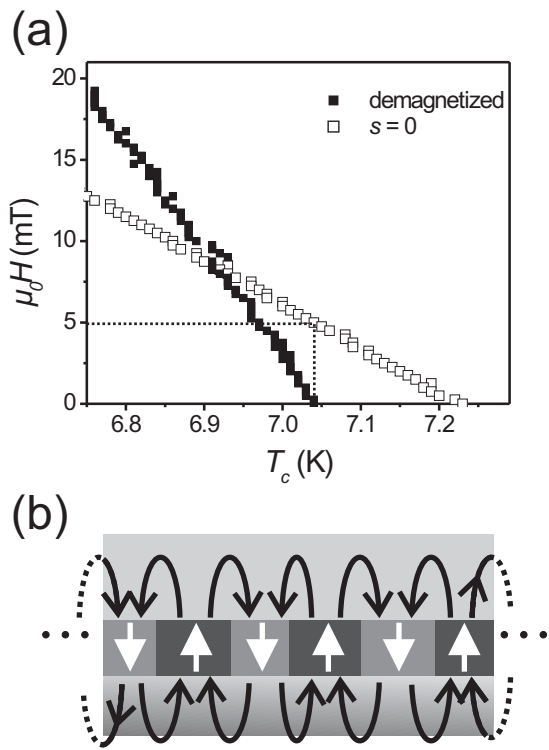}
\caption{(a)~Magnetic field - temperature phase diagram of the superconducting Pb film
covering the Co/Pt multilayer. The Co/Pt multilayer was demagnetized before measuring
$T_{c}(H)$. As a reference we have added the phase boundary of the $s=0$ state. The
dashed lines are guides to the eye. (b)~Schematic drawing of the stray field
$B_{stray}$ when the FM has been demagnetized.}
 \label{pbdem}
\end{figure}

The phase boundary for the demagnetized state, corresponding to the MFM image shown in
Fig.~\ref{moke}(b), is shown in Fig.~\ref{pbdem}(a).
In this state, $T_{c0}$ is suppressed to $(7.048 \pm 0.002)$~K. Moreover, the phase
boundary still shows a linear behavior, but with a slope increased by a factor of 2.4.
The difference between the phase boundaries in the demagnetized and the $s=0$ state
can be attributed to the influence of the stray field $B_{stray}$, suppressing the
order parameter in the superconductor above $T = 7.048$~K. The coherence length at
this temperature is $\xi=260$~nm. This means that superconductivity nucleates when the
value of $\xi$ becomes smaller than approximately the width of the band domains. The
nucleation first takes place in regions of the Pb film where the effective field in
the $z$-direction $\mu_0 H_{eff,z} = \mu_0 H_z + B_{stray,z}$ is minimum. The
confinement of these superconducting nuclei leads to the different $T_c(H)$ dependence
compared to the magnetized state. Aladyshkin {\em et al.} have very recently
calculated the $T_{c}(H)$ phase boundary of similar systems as the one that is
experimentally investigated here in the framework of the linearized Ginzburg-Landau
equation.\cite{aladyshkin03} They found, in agreement with our experimental result,
that the upper critical field of a superconducting film can have very unusual
temperature dependencies when a single domain wall or periodic domain structures are
present in a ferromagnetic film that is in contact with the superconductor. Based on
these considerations, we can conclude that the increased slope of the $T_c(H)$ curve
may be related to the specific domain pattern in the demagnetized state.\\ The MFM
image in Fig.~\ref{moke}(b) shows an equal contrast above all bright or dark domains,
indicating that $B_{stray}$ is rather homogeneous above the domains, and inhomogeneous
above the domain walls, which define some sharp corners. From calculations of the
upper critical field of mesoscopic superconducting structures, e.g., triangles or
squares, it is well known that the nucleation of superconductivity takes place first
in the corners of these structures. \cite{fomin98, schweigert99} Therefore, when
trying to calculate the $T_c(H)$ phase boundary in order to explain the increased
slope, one should take into account these corners formed by the domain walls. This
could be done by expanding the one-dimensional model used by Aladyshkin {\em et
al}\cite{aladyshkin03} to two dimensions.

\subsection{$T_c(H)$ with bubbles in the FM}
\begin{figure}
\includegraphics{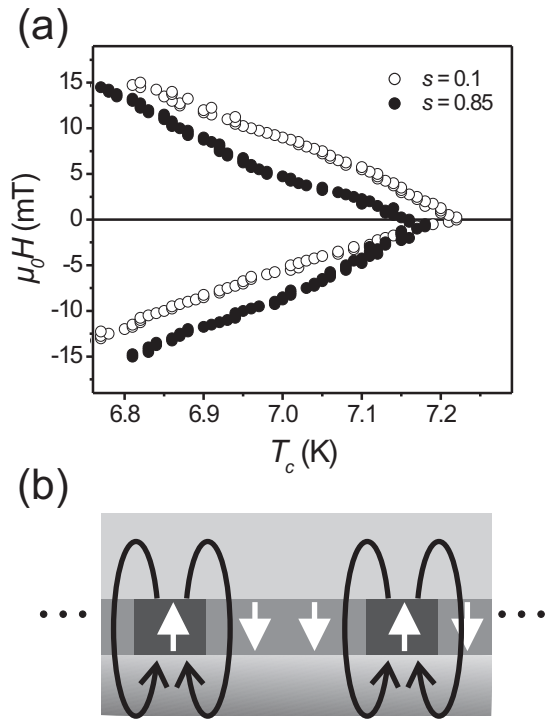}
\caption{(a)~Magnetic field - temperature phase diagrams of the superconducting Pb
film covering the Co/Pt multilayer. Before measuring $T_{c}(H)$, the Co/Pt multilayer
was brought into the $s=0.1$ and $s=0.85$ states.}
 \label{pbbub}
\end{figure}

\begin{figure}
\includegraphics{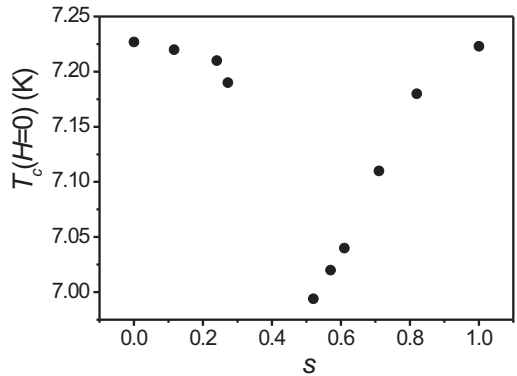}
\caption{Dependence of the critical temperature at zero field $T_{c}($H=0$)$ on the
parameter $s$. The minimum value of $T_c$ is observed for $s=0.5$.}
 \label{Tcs}
\end{figure}

The phase boundary shown in Fig.~\ref{pbbub}(a) is obtained when the Co/Pt multilayer
contains bubble domains.
Fig.~\ref{pbmag} and Fig.~\ref{pbdem} are symmetric with respect to $H$, i.e., $T_{c}$
is the same for positive or negative $H$, but the presence of the bubble domains
causes an {\it asymmetry of $T_{c}$ with respect to $H$}. For bubbles having positive
magnetic moments, i.e. $s<0.5$, a higher $T_{c}$ is observed for positive $H$ than for
corresponding negative $H$, whereas for bubbles containing negative magnetic moments
($s>0.5$), $T_{c}$ is higher for negative $H$. Moreover, both $T_c(H)$ curves shown in
Fig.~\ref{pbbub}(a) show a non-linear behavior with bumps in the field ranges around
$|\mu_0 H| \approx 5-10$~mT.\\ To explain the asymmetric $T_c(H)$ curves, let us
assume that the sample contains bubble domains with $m_z>0$ in a matrix of magnetic
moments with $m_z<0$, as shown in Fig.~\ref{pbbub}(b): $B_{stray,z}$ is positive above
the bubbles and negative between them. A positive $H$ in the $z$-direction compensates
the negative $B_{stray,z}$ between the bubbles and enhances $B_{stray,z}$ above them,
while a negative $H$ has the opposite effect: it enhances $B_{stray,z}$ between the
bubbles and compensates $B_{stray,z}$ above them. The important point that causes the
asymmetric phase boundary is that the absolute value of $B_{stray,z}$ is larger above
the $m_z>0$ regions (bubbles) compared to the $m_z<0$ regions (between the bubbles).
When the sample is cooled in positive $H$, superconductivity can nucleate at higher
temperatures in the area between the bubbles (where $B_{stray,z}<0$), compared to
cooling the sample in the corresponding negative $H$, where the nucleation takes place
in the areas above the bubbles. Note that qualitatively similar non-linear $T_c(H)$
curves as those presented in Fig.~\ref{pbbub}(a) have also been predicted by
Aladyshkin {\em et al}.\cite{aladyshkin03}\\ The critical temperature of the
superconductor decreases when the bubble domains have larger density. To illustrate
this effect, Fig.~\ref{Tcs} shows the dependence of $T_c(H=0)$ on the parameter $s$.
A clear minimum of $T_c$ is observed around $s=0.5$, which indicates that this domain
state has the largest value of the stray field of all investigated domain structures.
Note that $T_c$ of the $s=0.5$ state is even lower than $T_c$ of the demagnetized
state, emphasizing the inhomogeneous character of the $s=0.5$ state. The phase
boundary of this domain state will be discussed in the next section.

\subsection{$T_c(H)$ with the FM in the $s=0.5$ state}
\begin{figure}
\includegraphics{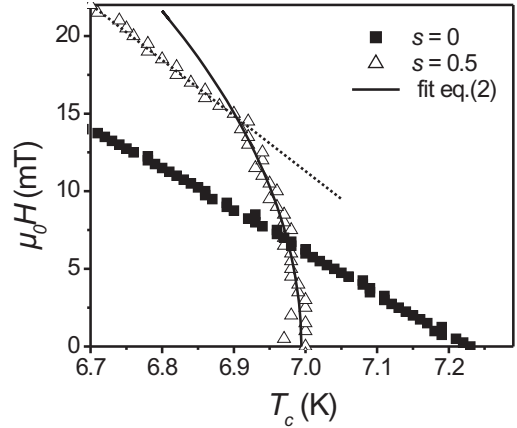}
\caption{Magnetic field - temperature phase diagrams of the superconducting Pb film
covering the Co/Pt multilayer. Before measuring $T_{c}(H)$, the Co/Pt multilayer was
brought into the $s=0.5$ state. As a reference the phase boundary of the $s=0$ state
is added. The dashed line is a guide to the eye.}
 \label{pb05}
\end{figure}

The phase boundary of the SC with the FM in the $s=0.5$ state is shown in
Fig.~\ref{pb05}.
In this domain state, $B_{stray}$ has a more inhomogeneous character than in the
demagnetized state. For a discussion of the differences between these two domain
states we refer to section~\ref{secmag}. $T_{c}(H)$ follows a non-linear behavior, in
contrast to the demagnetized and the $s=0$ states. $T_{c}(H)$ for the $s=0.5$ state
can not be described by eq.~\ref{one}, but rather by eq.~\ref{two} in fields $\mu_0
H<15$~mT, see the fit in Fig.~\ref{pb05}(b). This indicates that in the $s=0.5$ state
the regions where superconductivity nucleates can be considered as superconducting
strips with a width $w \leq \xi(T)$, forming a sort of a superconducting network. When
fitting the $T_{c}(H)$ curve using eq.~\ref{two} and $\xi(0)=41.2$~nm (from the phase
boundary for $s=0$), we obtain values of $w=(213 \pm 6)$~nm and $T_{c0}=(6.994 \pm
0.003)$~K. The value of $w$ determined from this fit can be compared with the typical
bubble domain size of $\sim 300$~nm. For $\mu_0 H>15$~mT and $T<6.90$~K, $T_{c}(H)$
shows a crossover from the one-dimensional network like to a two-dimensional linear
behavior, because the assumption $w < \xi(T)$ for eq.~\ref{two} is no longer
fulfilled. This is in agreement with the value of $\xi(6.90~\textrm{K})=194$~nm.
\section{Conclusion}
In conclusion, the phase boundary between the normal and the superconducting state of
FM/SC bilayers has been found to be strongly dependent on the domain structure in the
FM. The stray field $B_{stray}$ of these domains can lead to a significant decrease of
$T_{c}$ in zero applied field, but, on the other hand, it can also enhance $T_{c}$ in
applied fields. It has been demonstrated that the presence of bubble domains leads to
the formation of the field-polarity dependent asymmetric phase boundaries $T_c(H)$
with respect to $H$, due to compensation effects between $H$ and $B_{stray}$. For a
specific inhomogeneous domain structure, the $T_c(H)$ phase boundary shows a
crossover from a one-dimensional to a two-dimensional nucleation behavior.\\
\begin{acknowledgments}
The authors thank L. Van Look, K. Temst and G. G\"untherodt for help with sample
preparation, J. Swerts for MOKE measurements and Y. Bruynseraede for fruitful
discussions. This work was supported by the Fund for Scientific Research - Flanders
(Belgium) (F.W.O.-Vlaanderen), by the Belgian IUAP and the European ESF "VORTEX"
Programs, and by the Research Fund K.U.Leuven GOA. ML and MJVB are Postdoctoral
Research Fellows of the F.W.O.-Vlaanderen.
\end{acknowledgments}

\end{document}